\documentclass[12pt,preprint]{aastex}
\usepackage{graphicx}
\begin{document}

\slugcomment{Accepted in Astrophysical Journal Letters}

\title{Molecular Signatures in the Near Infrared Dayside Spectrum of HD 189733b}

\author{M.~R. Swain\altaffilmark{1},
G. Vasisht\altaffilmark{1}, 
G. Tinetti\altaffilmark{2},
J. Bouwman\altaffilmark{3},
Pin Chen\altaffilmark{1},
Y. Yung\altaffilmark{4},
D. Deming\altaffilmark{5}, \&
P. Deroo\altaffilmark{1}}

\altaffiltext{1}{Jet Propulsion Laboratory California, Institute of Technology, 4800 Oak Grove Drive,
Pasadena, CA 91109} 
\altaffiltext{2}{University College London, Gower Street, London WC1E 6BT, UK}
\altaffiltext{3}{Max-Planck Institute for Astronomy, Konigstuhl 17,
D-69117 Heidelberg, Germany}
\altaffiltext{4}{Division of Geological and Planetary Sciences, California Institute of Technology, Pasadena, CA 91125}
\altaffiltext{5}{Goddard Space Flight Center, Planetary Systems
Branch, Code 693, Greenbelt, MD}

\begin{abstract}

We have measured the dayside spectrum of HD 189733b between 1.5 and
2.5 $\mu$m using the NICMOS instrument on the Hubble Space Telescope.
The emergent spectrum contains significant modulation, which we
attribute to the presence of molecular bands seen in absorption.  We
find that water (H$_{2}$O), carbon monoxide (CO), and carbon dioxide
(CO$_{2}$) are needed to explain the observations, and we are able to
estimate the mixing ratios for these molecules.  We also find
temperature decreases with altitude in the $\sim 0.01<P<$ $\sim 1$ bar
region of the dayside near-infrared photosphere and set an upper limit
to the dayside abundance of methane (CH$_{4}$) at these pressures.

\end{abstract}
\keywords{planetary systems --- techniques: spectroscopic}

\section{Introduction}

HD~189733b is a transiting hot-Jupiter planet in a 2.2-day orbit
around a K2V stellar primary \citep{bouchy05}. Due to the relatively
large depth of the eclipse ($\sim2.5$\% at K band) and the bright
stellar primary (Kmag = 5.5), this system was immediately recognized
as an important target for atmospheric characterization observations,
and its emission has been studied extensively durning secondary eclipse
\citep{deming05,knutson07,grillmair07,knutson08,charbonneau08}.
Multicolor photometric observations revealed the presence of H$_{2}$O
\citep{tinetti07a} and the likely presence of CO
\citep{charbonneau08}, while optical transmission spectroscopy
suggests scattering by high altitude haze \citep{pont08}.  In
addition, extensive theoretical work has also been done on the
atmosphere of this planet by
\cite{fortney06,barman08,burrows08,showman08} and others.

Recently, the NICMOS camera on the Hubble Space Telescope (HST) was
used during the primary eclipse of HD~189733b to obtain a near-IR
transmission spectrum of the planet's atmosphere; these results showed
the presence of H$_{2}$O and CH$_{4}$ (Swain et al. 2008; hereafter
SVT08).  The near-IR transmission spectrum of HD~189733b probes the
upper (P $\sim 10^{-4}$ bar estimated from our models) regions of the
of the atmosphere at the terminator.  In this paper, we report the
results of HST observations of the dayside spectrum (derived from
secondary eclipse measurements).  At near infrared wavelengths, the
dayside portion of the atmosphere of HD~189433b is probed in deeper (P
$\sim 0.1$ bar estimated from our models) regions.

\section{Observations}
\label{observations}

We observed HD~189733 with the Hubble Space Telescope for five
contiguous spacecraft orbits, using NICMOS in imaging-spectroscopy
mode with the G206 grism (wavelength coverage 1.4-2.5 $\mu$m).
Observations began on April 29, 2007 at UT 23:47:58 and ended at UT
06:51:52 on the next day. The first two orbits ($O_1$ \& $O_2$)
observed the target pre-ingress, the third orbit ($O_3$) was phased to
capture the occultation, while the fourth and fifth orbits ($O_4$ \&
$O_5$) were post-egress (see figure 1).  A total of 638 usable
snapshot spectra were acquired during the five orbits. The third orbit
(the occultation) contained 130 spectra. Using the best available
ephemerides \citep{winn07,pont08} we determine that full-occultation starts
$\sim 13$ exposures into $O_3$, and lasts for the remainder of this
orbit. The effective exposure time for each spectrum was $T = 1.624$
s; the overall instrumental configuration, including the location of
the spectrum on the focal plane array, was identical to that in
SVT08. A few spectral calibration exposures (in $O_1$) were acquired
using a narrow-band filter.

Because the NIC-3 camera is severely undersampled and because of large
gain drop-off at the edges in the NICMOS detector pixels, the
instrument was configured in DEFOCUS mode (FWHM $\simeq$ 5
pixels). This defocus reduces the level of photometric fluctuations
due to both pointing jitter (random) and beam wander (systematic
errors). While the defocus also helps to minimize the bright-source
overheads for this extremely bright target, it nevertheless limits the
effective spectral resolution to $R \simeq 40$.

\subsection{Data Analysis}

A complete description of the data-analysis methods is provided in the
supplementary information to SVT08. Herein, we discuss only departures
and addenda to these analysis methods.

First, each image is reduced to a one-dimensional spectrum covering
1.5-2.5 $\mu$m in spectral range. This includes combining data from
the grating's first and second orders, which are simultaneously imaged
onto the detector array. Inclusion of the second order improves the
SNR in the blue because the second order, which is partially imaged
onto the detector, contains more detected photon-flux at the shortest
wavelengths (shortwards of 1.7 $\mu$m, longwards of which the
efficiency of the 2$^{nd}$ order falls rapidly) as a consequence of
the peculiar blaze of the parent grating.  Prior to combination,
wavelength solutions for both orders were obtained using the
calibration exposures and the wavelength-scale coefficients in
Nicmoslook \citep{freudling97}.  Merger of the spectral orders takes
advantage of the near integer ratio, $\simeq 2$, of the spectral
resolutions of the two orders.  The orders are coadded around a
fiducial wavelength, and any smearing is captured by assigning the
coadded channel a wavelength equivalent to the weighted average of
wavelengths of component channels. Prior to coadding, the 2$^{nd}$
order spectra were multiplied by a broad Tukey window so as to
suppress the effects of (1) the sharp band-edge at 1.5 $\mu$m
and (2) the detector array edge at $\sim 1.85$ $\mu$m on the final
spectrophotometry.  Lastly, coadded spectra are rebinned to
wavelength channels at the binwidth of the DEFOCUS
full-width-at-half-maximum, guaranteeing adjacent channels have
independent spectral content. As the state variables are channel
independent, the two orders can be coadded prior to decorrelation (see
below).  The secondary eclipse is observable in the raw light-curves
(see Fig. 1) and especially so in spectral subbands with high contrast
({\it e.g.} around 2.3 $\mu$m; see Fig. 3).  For example, if we divide
the ``red broadband'' (see caption Fig. 1) into two subbands between
1.9 to 2.2 $\mu$m and 2.2 to 2.5 $\mu$m, little contrast is apparent
in the former while a large contrast is seen in the latter.

The nature of the state variables chosen for decorrelation has been
discussed extensively in SVT08-SI. One change is in the array
temperature vector: instead of using temperatures from the probe at
the NIC-1 mounting cup, we use the diode properties of the detector
pixels as proxy for the detector temperature because the measurable
detector bias voltage is highly temperature sensitive. These
temperatures were extracted by the STScI staff on request (courtesy
N. Pirzkal, priv. comm.).  We find that the peak-to-valley variations
of the state variables are similar (i.e.  $< 0.1$ pixel for PSF
displacement) to those seen in the observations of SVT08 for all
orbits included in the fitting procedure, i.e. $O_{2}-O_{5}$. A large
rotation of the dispersion axis is apparent in $O_1$, and $O_1$ data
were discarded in order to satisfy our underlying assumption -- that
spectrophotometry systematics are linear in small perturbations of the
state variables. After $O_1$ the rotation settles, although $O_2$
shows a larger rotation from mean than the final three orbits.

The framework for decorrelation of the light curves is the joint
energy minimization procedure described in SVT08, i.e. modelling and
removal of the systematics was performed simultaneously, with the
eclipse-depth, $\alpha(\lambda)$, left as a free variable.  The unique
$\chi^2$ minima in the $\alpha-\lambda$ space determines the emission
spectrum. The point-to-point photometric scatter (see Fig. 1), which
is stochastic in time but common to all wavelength channels, is
estimated using optimally weighted fit-residuals as in SVT08-SI. For
all channels, the final residual vectors are examined via Lomb-Scargle
periodograms to ensure that correlated $1/f$ noise, as well as noise
at the HST orbital period and its overtones, is removed.  An
alternative scheme, that uses just the baseline light curves to fit for
a model (SVT08-SI) and then interpolates to determine the emission
spectrum, gave answers that were statistically consistent with the
joint energy minimization scheme; this implies that just the baseline
orbits provide an adequate model.

The importance of the various state variables in modelling the overall
systematics is assessed via principal components.  The model in each
light curve is first decomposed along each state vector, which is
converted to a channel-wise correlation matrix. Each correlation
matrix is diagonalized, along with the computation of the principal
vectors.  We find that a maximum of two principal components is
required to model the bulk of the variance in the channel
light curves. The rotation $\hat\theta$, and the displacements $\hat
x$, $\hat y$, are the most important projections onto the first
principal component, which models between 55-90 \% of the variance
across channels. The spacecraft orbital phase, $\hat\phi_H$, is the
single most important projection onto the second principal
component. This behavior is largely explained by the fact that most
elements in the set of state vectors show some dependence on the
spacecraft orbital period.

The final emission spectrum (planet/star ratio) is shown in figures 2
\& 3 and tabulated in Tab. 1. The error bars contain three major
components in quadrature.  These are as follows: (1) the random noise,
including detection noise, errors in the estimation of the state
vectors, and any remaining systematics are contained in the light
curve residuals, (2) stochastic uncertainties in the model are
computed by bootstrapping, and (3) gross errors in the model
fitting. The latter are estimated either via Monte Carlo methods by
applying random perturbations to the model components (SVT08-SI), or
via the $\chi^2$ surfaces obtained from the fitting procedure.
Finally, we draw attention to the fact that the penultimate bluest
channel (1.525 $\mu$m) has an unphysical eclipse depth; specifically,
the wavelenth channel is negative at $2.5\sigma$.  This may be due to
the fact that this channel contains a dead pixel in the brightest part
of the spectral image. We have not evaluated the effect of such
dead-pixels on the spectrophotometry in a strict statistical sense;
consequently such effects are not reflected in the error-bar for that
channel.  Previous ground-based observations have been used to
estimate the K-band planet/star contrast upper limit as 0.00039
\citep{barnes07}.  We find the contrast in the same passband
(averaging the contrast in our spectral channels between 1.98 and 2.38
$\mu$m) to be 0.0004; given the challenges associated with
ground-based secondary eclipse measurements, we do not consider the
difference significant.

\section{Discussion}

The interpretation of emergent spectra is generally
complex because molecular bands could appear either in emission,
absorption, or both, depending on the detailed temperature, pressure,
and chemical profile of the planetary atmosphere. Nonetheless, when
molecular species with detectable spectral features are present,
radiative transfer models can be used to retrieve the temperature and
abundance structure of the photosphere of emission. This approach,
used extensively for the study of planetary atmospheres in our solar
system \citep{goody89}, is the basis of our spectral retrieval.

We undertook a detailed analysis using a line-by-line,
radiative-transfer model developed for disk-averaged planetary spectra
and subsequently adapted for the specific case of hot-Jupiters
\citep{tinetti06, tinetti07b}. The model covers a pressure range from
$1 \le P \le 10^{-6}$ bar in which the individual model layers are
populated by abundance profiles for molecules used in the retrieval
process. The opacity contribution of each molecule is computed based
on its mixing ratio $c_i$, local density ($\rho$), and temperature in
accordance with the assumed T-P profile. Our compendium of line-lists
is the same as that given in SVT08. Trial planetary spectra were
converted to contrast spectra using an ATLAS-9 model-spectrum of a
K1-K2V star with solar abundances \citep{kurucz93}. To retrieve
molecular abundances and the temperature profile, we iteratively
compared the modelled contrast spectra to the observed NIR spectra as
well as mid-IR Spitzer-IRAC photometry \citep{charbonneau08}.  We
found that the IRAC photometry did not provide constraints for
molecular mixing ratios, although those data were consistent with a
decreasing T profile.  Thus, the results for molecular abundances
presented here are independent of IRAC data that probe lower
pressures in the atmosphere.

The significant spectral modulation present in the near-IR contrast
spectrum is best explained by the presence of water vapor and the
carbon-bearing molecules CO and CO$_2$. The model requires a
decreasing temperature with altitude between $\simeq$ 1 $<$ P $<$
$10^{-2}$ bar (the relevant photosphere).  The modelling began by
including the minimum number of molecules (H$_{2}$O, CO, and CH$_{4}$)
composed from the most abundant reactive elements (H,O,and C).  These
molecules are highly plausible; H$_2$O and CH$_4$ have already been
inferred via transmission photometry and spectroscopy
\citep{tinetti07a,swain08b}, and CO, thermochemically the most stable
carbon molecule on the hot day-side, has been inferred from photometry
\citep{barman08,charbonneau08}.  However, this minimal combination of
molecules is unable to reproduce the large contrast deficit in the
1.95-2.15 $\mu$m region of the spectrum, suggesting additional opacity
by the 2 $\mu$m combination-bands of CO$_2$. The inclusion of CO$_2$
in the model reproduces the major features of the observed
spectrum. Fig. 2 illustrates the contribution of individual molecules
to the model spectrum; the significance of CO$_2$ is evident.  This
suggests searching for evidence of other CO$_{2}$ bands, such as the
15 $\mu$m band, which may be present in the Spitzer mid-IR spectra
\citep{grillmair07}.

Although CO$_{2}$ is not the primary carrier of carbon in the pressure
and temperature range of relevance here, it is thermochemically
predicted to be fairly abundant ($10^{-6}$ for solar abundance) when
CO is the major carbon-bearing gas \citep{lodders02} and has been
discussed in the context of low-gravity brown dwarf spectral models
\citep{saumon03,mainzer07}.  It is formed via the net thermochemical
reaction,
$$ \hbox{\rm CO}~+~\hbox{\rm H$_2$O} \rightleftharpoons \hbox{\rm
CO$_2$}~+~\hbox{\rm H$_2$}.$$ Additionally, CO$_{2}$ is readily
produced by the reaction $$ \hbox{\rm CO}~+~\hbox{\rm OH}
\rightleftharpoons \hbox{\rm CO$_2$}~+~\hbox{\rm H}.$$ where the
hydroxyl radical OH is derived from photolysis of H$_{2}$O
\citep{liang03}.

While the observed spectrum places strong constraints on what major
molecules are present, the model fitting must deal with a degeneracy
between $dT/dP$ and the molecular mixing ratio. In order to estimate
the range of plausible mixing ratios for the inferred species, our
final family of models explored a range of $T$ profiles
\citep{barman08,burrows08}, with constant vertical mixing ratios for
H$_2$O, CO and CO$_2$, and a vertical mixing profile for CH$_4$, as
derived by Liang et al. 2003. Based on this family of models, we find
the following range for mixing ratios at pressure altitudes relevant
here: H$_{2}$O ($c \sim 0.1-1 \times 10^{-4}$), CO ($c \sim 1-3 \times
10^{-4}$), CO$_{2}$ ($c \sim 0.1-1 \times 10^{-6}$), and CH$_{4}$ ($c
\leq 1 \times 10^{-7}$).  These values imply a 0.5 $\leq$ C/O $\leq$
1 together with a metallicity that is potentially subsolar.  The
inclusion of CO$_2$ in our models does not explain the contrast
deficits at the band-edges, particularly the blue edge at 1.6 $\mu$m
(Fig. 3). One possibility is the presence in the atmosphere of trace
amounts of hydrocarbons such as acetylene (C$_2$H$_2$), ethane
(C$_2$H$_6$), and possibly ammonia (NH$_3$).  For example, C$_2$H$_2$,
CH$_4$ and C$_2$H$_6$ can be produced via photolysis of CO (the
primary reservoir for carbon) and water \citep{liang04}, while ammonia
is likely to be found in low concentrations (c $\simeq 10^{-7}$) in
thermal equilibrium with N$_2$, which is the dominant reservoir for
nitrogen in these conditions. Indeed, including one or more of the
above molecules does improve the model fit at the band-edges.
However, a firm assertion of their presence amongst a myriad of
possibilities would require better data and/or broader wavelength
coverage.

\section{Conclusions}

In summary, we have presented the first near infrared spectrum of
light emitted by a hot-Jupiter type planet. Using a radiative transfer
model, we determine that the molecules H$_{2}$O, CO and CO$_{2}$ are
likely present on the dayside of HD~189733b, and we are able to
estimate abundances for these species.  Although we cannot tightly
constrain the slope of the temperature profile, the observed
absorption features indicate that temperature decreases with altitude
at pressures between 1 and 0.01 bar.  In addition, there are residual
features in the spectrum that could be explained by including
opacities of other species that are presently not well constrained.

This work represents an initial step in exploiting the power of
spectral analysis (even at low resolutions of $R \simeq 40$) in
determining the chemical composition of extrasolar planetary
atmospheres; it also illustrates the extraordinary potential of the
NICMOS instrument for characterizing bright, transiting exoplanets.
In a previous paper, we presented a transmission spectrum of this
planet at the terminator (SVT08), in which methane is seen to be more
abundant ($c$ = $5\times10^{-5}$) in the higher, cooler regions of the
terminator region atmosphere.  However, it is difficult to compare
directly the previous terminator results with our current dayside
results because they probe atmospheric regions with significantly
different temperatures and altitudes on this highly irradiated
planet. The development of sophisticated global circulation and
chemistry models could significantly advance the state of the art in
this respect.

\acknowledgments

We thank Tommy Wiklind, Nor Prizkal, and other members of the Space
Telescope Science Institute staff for extensive assistance in planning
the observations and for providing advice about ways in which the
observations could be optimized.  We also thank Jonathan Fortney for
helpful recommendations on improving the presentation of this material.
G. Tinetti was supported by the UK Sciences \& Technology Facilities
Council.  A portion of the research described in this paper was
carried out at the Jet Propulsion Laboratory, under a contact with the
National Aeronautics and Space Administration.

\begin{figure}[h!]
\begin{center}
\epsscale{0.5}
\includegraphics[angle=0,scale=0.7]{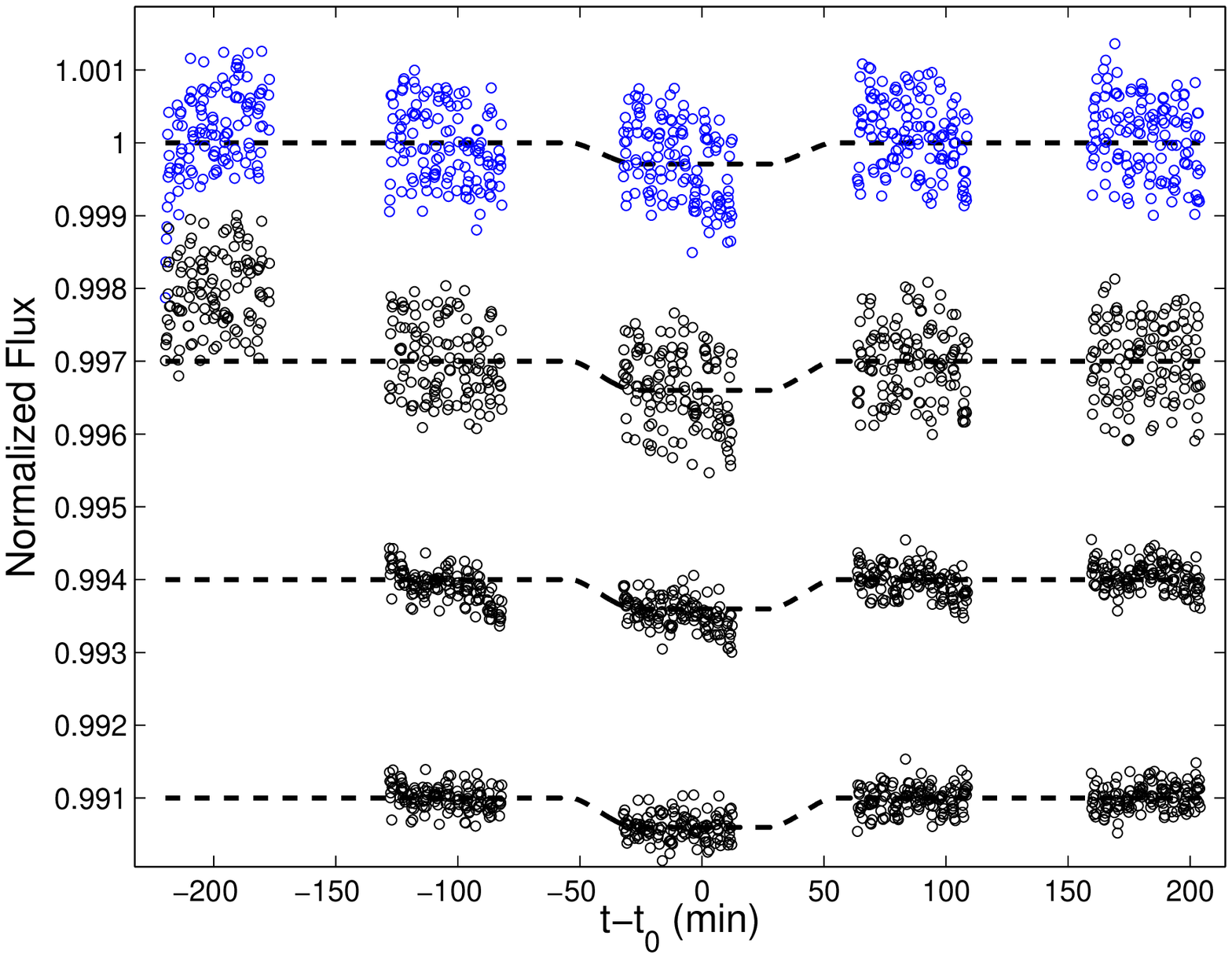}
\caption{This figure shows a set of normalized light curves plotted
against time (in minutes; $t_0$ corresponds to MJD 54220.1442888) in a
top-to-bottom sequence.  To aid the eye, we plot the data together with
a set of model light curves calculated assuming a uniform-disk for HD~189733b
passing behind the parent star. (A) The top curve is a raw broadband or
``white-light'' light curve showing all five orbits.  The small
broadband eclipse depth of $2.9 \times 10^{-4}$ is barely obvious in
this light curve. (B) Next is a raw ``red-broadband'' light curve covering
a wavelength range of 1.9-2.44 $\mu$m. The somewhat larger
photometric depth of $4.1\times 10^{-4}$ is more easily discernible
here. Orbit 1 data are poorly behaved over this restricted band. (C)
Next is the ``red-broadband'' light curve after the removal of
point-to-point photometric correlations. This step tightens the
scatter in the light curve. Note that temporal correlations are not
removed and are clearly seen in the light curve. (D) At the bottom is a
fully processed light curve with all correlations modelled and removed.
Gaps in the time series occur because HD~189733 is not in the
continuous viewing zone for the Hubble Space Telescope.
\label{fig:lightCurve}}
\end{center}
\end{figure}

\begin{figure}[h!]
\begin{center}
\epsscale{0.5}
\includegraphics[angle=0,scale=0.7]{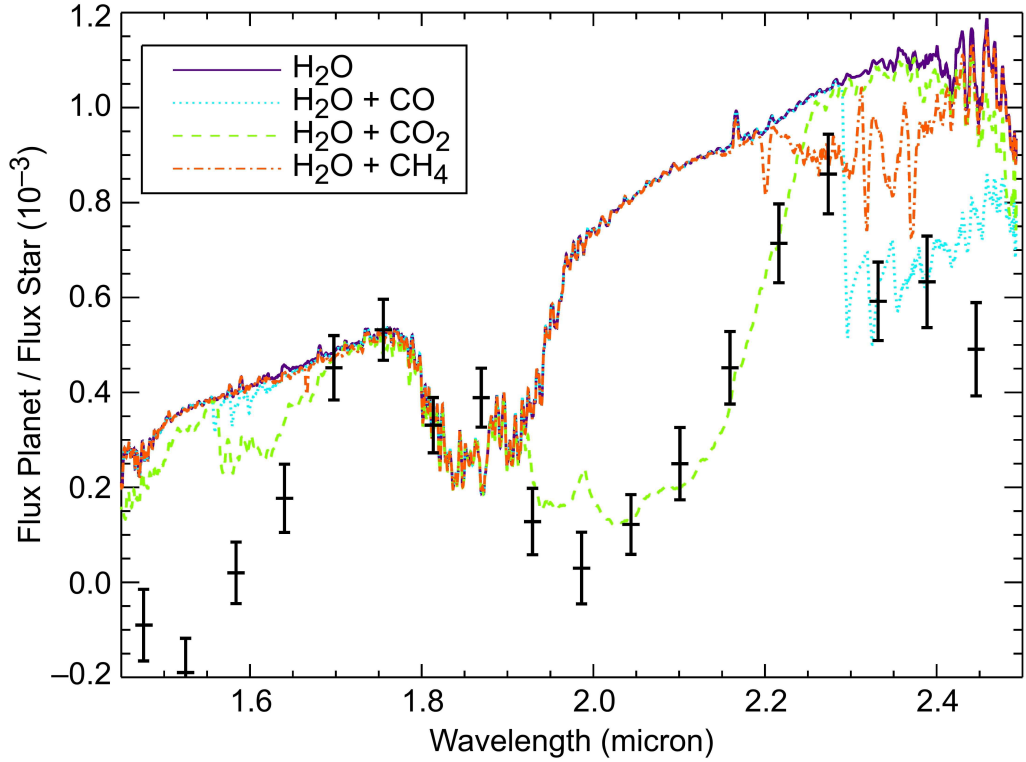}
\caption{ A comparison of the observed spectrum (black markers with
error bars) to a sequence of component models showing the affect of
water together with individual carbon-bearing molecules included in our final
model.  This comparison shows the portion of the spectrum where each
molecule has significant opacity.  In particular, the role of
CO$_{2}$ in producing the observed minima between 1.95 and 2.15 $\mu$m
is evident.
\label{fig:molecules}}
\end{center}
\end{figure}

\begin{figure}[h!]
\begin{center}
\epsscale{0.5}
\includegraphics[angle=0,scale=0.7]{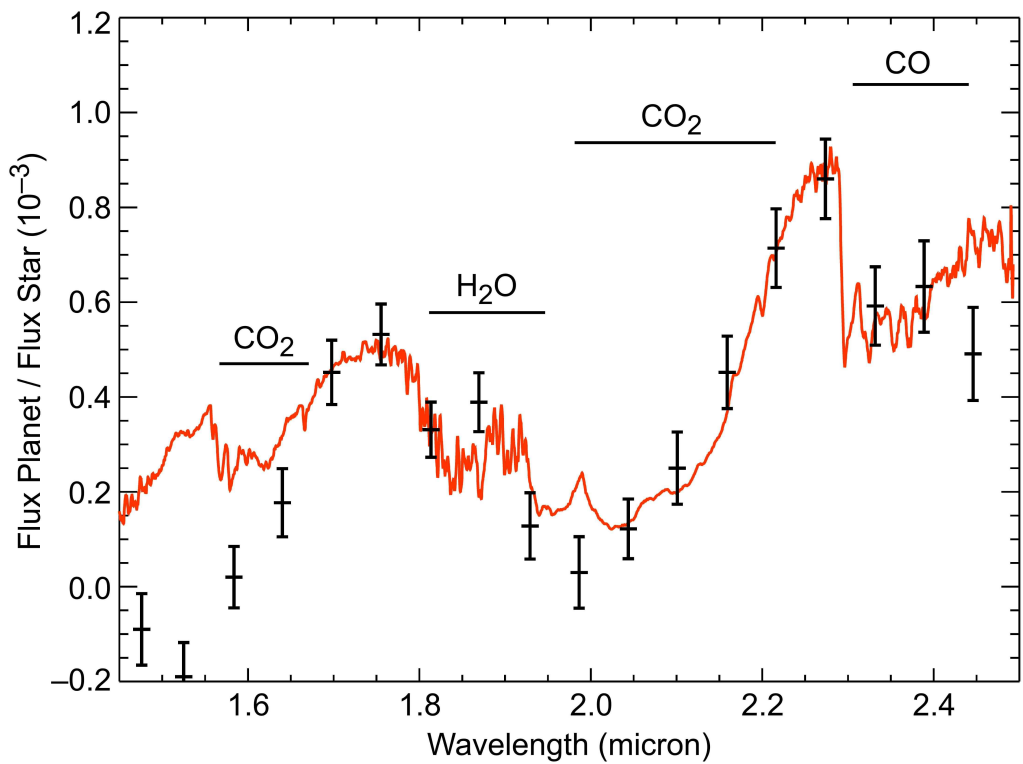}
\caption{The near-IR dayside emergent spectrum used in our analysis
  with $\pm 1-\sigma$ errors shown (black), together with a model
  spectrum (orange) containing the molecules H$_{2}$O, CO, CH$_{4}$
  and CO$_{2}$, which are responsible for the absorption features (the
  strongest of which are identified above).  The model also includes
  H$_{2}-$H$_{2}$, which contributes continuum opacity.  The fit
  residuals suggest that one or more additional molecular species may
  be present.  Although the fit is improved slightly by including
  C$_{2}$H$_{2}$, C$_{2}$H$_{6}$, or NH$_{3}$, additional data is
  required to make a strong case for the presence of additional
  molecular species.}
\end{center}
\end{figure}

\begin{table}[h]
\begin{center}
Planet/Star Contrast Spectrum \\
\begin{tabular}{|c|c|c|} \hline\hline
wavelength ($\mu$) & $F_{planet}/F_{star}$ & error \\ \hline \hline
 2.446 &  0.000491 &  0.00009 \\
 2.389 &  0.000633 &  0.00009 \\
 2.332 &  0.000592 &  0.00008 \\
 2.274 &  0.000860 &  0.00008 \\
 2.216 &  0.000714 &  0.00008 \\
 2.159 &  0.000452 &  0.00007 \\
 2.101 &  0.000250 &  0.00007 \\
 2.044 &  0.000122 &  0.00006 \\
 1.986 &  0.000030 &  0.00007 \\
 1.929 &  0.000128 &  0.00007 \\
 1.869 &  0.000389 &  0.00006 \\
 1.814 &  0.000331 &  0.00006 \\
 1.755 &  0.000532 &  0.00006 \\
 1.698 &  0.000452 &  0.00007 \\
 1.640 &  0.000177 &  0.00007 \\
 1.584 &  0.000021 &  0.00006 \\
 1.525 & -0.000190 &  0.00007 \\
 1.476 & -0.000089 &  0.00008 \\ \hline
\end{tabular}
\label{tab:spectrum}
\caption{Data and 1-$\sigma$ error for the contrast
  $F_{planet}/F_{star}$ spectrum of the dayside of HD~189733b.}
\end{center}
\end{table}

\end{document}